\documentclass[10pt,letterpaper]{article}
\usepackage[top=0.85in,left=2.75in,footskip=0.75in,marginparwidth=2in]{geometry}

\usepackage[utf8]{inputenc}

\usepackage{cite}
\usepackage{amssymb}
\usepackage{nameref,hyperref}

\usepackage[right]{lineno}

\usepackage{microtype}
\DisableLigatures[f]{encoding = *, family = * }

\raggedright
\usepackage{changepage}

\usepackage[aboveskip=1pt,labelfont=bf,labelsep=period,singlelinecheck=off]{caption}

\makeatletter
\renewcommand{\@biblabel}[1]{\quad#1.}
\makeatother

\usepackage{lastpage,fancyhdr,graphicx}
\usepackage{epstopdf}
\fancyheadoffset[L]{2.25in}
\fancyfootoffset[L]{2.25in}

\usepackage{color}

\definecolor{Gray}{gray}{.25}

\usepackage{graphicx}

\usepackage{sidecap}

\usepackage{wrapfig}
\usepackage[pscoord]{eso-pic}
\usepackage[fulladjust]{marginnote}
\reversemarginpar

\begin{document}
\vspace*{0.35in}

% title goes here:
\begin{flushleft}
{\Large
\textbf\newline{Side-polished Silica-Fluoride Multimode Fibre Pump Combiner for Mid-IR Fibre Lasers and Amplifiers}
}
\newline
% authors go here:
\\
Boris Perminov\textsuperscript{1},
Kirill Grebnev\textsuperscript{1},
Uwe H{\"u}bner\textsuperscript{1},
Maria Chernysheva\textsuperscript{1},

%\\
\bigskip
\bf{1} Leibniz Institute of Photonic Technology, Leibniz-IPHT, Albert-Einstein-Stra{\ss}e 9, 07745 Jena, Germany
\\
\bigskip
* Boris.Perminov@leibniz-ipht.de

\end{flushleft}

\begin{abstract}
Side-pumping fibre combiners offer several advantages in fibre laser design, including distributed pump absorption, reduced heat load, and improved flexibility and reliability. These benefits are particularly important for all-fibre lasers and amplifiers operating in the mid-IR wavelength range and based on soft-glass optical fibres. However, conventional fabrication methods face limitations due to significant differences in the thermal properties of pump-delivering silica fibres and signal-guiding fluoride-based fibres. To address these challenges, this work introduces a design for a fuse-less side-polished (D-shaped) fibre-based pump combiner comprising multimode silica and double-clad fluoride-based fibres. The results demonstrate stable coupling efficiency exceeding  80\% at a 980-nm wavelength over 8 hours of continuous operation under active thermal control. The developed pump combiner has also been successfully integrated into a linear Er-doped fibre laser cavity, showing continuous-wave generation at 2731 or 2781-nm central wavelength with an output power of 0.87~W. Overall, this innovative approach presents a simple, repeatable, and reproducible pump combiner design that opens up new possibilities for leveraging fibre-based component technology in soft glass matrices and other emerging fibres with unique compositions.   

\end{abstract}

%\begin{document}
%\maketitle

\section*{Introduction}
\noindent

In the 2000s, a notable increase in research and industrial attention towards mid-IR light sources and sensors stemmed from the rapid expansion of diverse industrial sectors. The raised emphasis on developing Mid-IR sources, specifically within the range of 2.7 to 4.5~$\mu$m, has been propelled by the rising need for applications such as monitoring greenhouse gases and pollutants~\cite{scherer2013mid}, optical frequency standards for global positioning systems and optical clocks~\cite{gubin2009femtosecond}, free space and fibre optical communications~\cite{zlatanovic2010mid}, LIDAR systems, and medical applications, particularly for spectroscopic diagnostics ~\cite{baker2014using,seddon2013mid}. As a result, over the past decade, fibre-based Mid-IR lasers have emerged as promising high-brightness light sources capable of generating light beyond 2.5~$\mu$m. The configuration of fibre lasers offers a tailored and flexible design in terms of output power and operational regime, as well as advantages of reduced cost, improved efficiency, and ease of operation compared to established quantum cascade laser technology.

To leverage the potential of fibre lasers in the mid-IR wavelength range, it is necessary to integrate unique soft-glass fibres into the existing infrastructure currently dominated by silica fibres. Silica fibres are unsuitable for mid-IR applications due to the high phonon energy of the glass matrix (1300~cm$^{-1}$). Such a high phonon energy leads to extreme absorption over 3~dB/m beyond 2.5~$\mu$m, which cannot be compensated by a small-signal gain in active fibre to initiate laser generation. On the contrary, fibres based on soft-glass matrices, such as fluoride or chalcogenide, feature low phonon energy, making them a promising platform for mid-IR system development. Particularly, the phonon energy of fluoride fibres is approximately $\sim600~cm^{-1}$, which enables transmission within the wavelength range from visible light up to approximately 5~$\mu$m~\cite{jackson2021mid,tao2015infrared}. Due to their soft glass nature, fluoride-based fibres have lower melting temperatures (280-400~$^{\circ}$C), higher coefficient of thermal expansion (179$\times$10$^{-7}$~K$^{-1}$) and lower Young's modulus ($\sim$60~GPa) when compared to harder glasses like silica (1300~$^{\circ}$C, 5.9$\times$10$^{-7}$~K$^{-1}$ and $\sim$70~GPa, correspondingly) \cite{shelby2020introduction,wang2009review}. Recent research results have demonstrated promising progress in the development of Mid-IR lasers based on fluoride fibres doped with Erbium (Er), Dysprosium (Dy), and Holmium (Ho)-doped fluoride fibres, primarily with ZBLAN (ZrF$_4$-BaF$_2$-LaF$_3$-AlF$_3$-NaF) or InF$_3$-based glass matrices, as a gain media for direct generation from 2.7 to 4.4$\mu$m~\cite{duval2015femtosecond,bawden2021ultrafast,majewski2018dysprosium,maes2018room,jackson2012towards,fortin2016watt}.

Yet, replacing silica fibres with soft-glass ones or even their combination within the laser setup sets challenging requirements of redesigning existing fibre post-processing methodologies~\cite{xia2023fabrication,seguin2023fabrication} or developing innovative concepts for essential fluoride fibre-based laser components~\cite{fernandez2022ultrafast}. One of the key examples is the design of effective pump combiners, where pump lasers are pigtailed with silica fibres. Conventionally, this development has been successfully done using the thermal treatment of silica fibres, i.e. tapering and fusion splicing~\cite{dimmick1999carbon,kosterin2004tapered,wang2009review}. Other approaches, such as various side-coupling techniques~\cite{huang2012direct,koplow2003new,xu2003non}, while demonstrating high efficiency in scientific results, have seen limited application in commercial systems due to their intricate manufacturing processes, higher complexity and cost. As mentioned above, the melting temperatures, thermal expansion coefficient and viscosity of fluoride and silica fibres are drastically different, which restricts fuse-based methodologies for the fabrication of fused hybrid fibre pump combiners.

In the quest to enhance the pumping arrangement in Mid-IR fibre lasers, achieve a uniform distribution of pump power along fluoride fibres, and avoid heat degradation of the fibre tip during face pumping~\cite{aydin2019endcapping}, a few designs of side pump combiners have previously been proposed. Namely, \textit{Sch{\"a}fer et al.} fabricated side combiner by splicing the angle-polished multimode fibre onto the first cladding of a double-clad fibre~\cite{schafer2018fluoride}. The pump combiner included two fluoride fibres that were used for both power delivery and signal generation. The coupling efficiency reached 83\%. Alternatively, a pump combiner based on fluoride fibres has been fabricated using the conventional heat-pull method, reaching nearly 50\% efficiency \cite{annunziato2022fused}. However, neither of these methodologies has provided an interface between pump-delivering silica fibre and fluoride signal fibre. In both works, the pump light from a pump source was coupled free-space into fluoride fibres~\cite{uehara2019power}.

Ultimately, the other class of pump combiners present hybrid designs comprising silica and fluoride fibres. Notably, the approach presented by \textit{Magnan-Saucier et al.} utilised core-less silica fibre tapered down to 12~$\mu$m waist diameter for the pump delivery~\cite{magnan2020fuseless}. The pump delivery taper was placed along the fluoride double-clad fibre, ensuring physical contact via surface tension and, hence, the record-high coupling efficiency of 93\%. The physical contact of the fibres is very critical as it can significantly limit the coupling efficiency. Therefore, such methodology employs quite a challenging task of twisting the fragile ultra-thin tapered and fluoride fibres. An analogous technology of tapering the pump-delivering silica fibre, yet only down to $\sim$60~$\mu$m in diameter, has been shown in \cite{Karampour2023all}. The tapered silica fibre has been fixed to the edge of the first cladding of double-cad ZBLAN fibre at a 9-degree angle using UV curable glue. As a result, the coupling efficiency of such a combiner mounted to 60\%.

Most recently, these two approaches have been combined to design a tellurite fibre beam combiner by fusion splicing tapered delivery fibre onto signal one~\cite{xia2023fabrication}. 

\begin{figure}[!bp]
    \centering
    \includegraphics[width=0.5\linewidth]{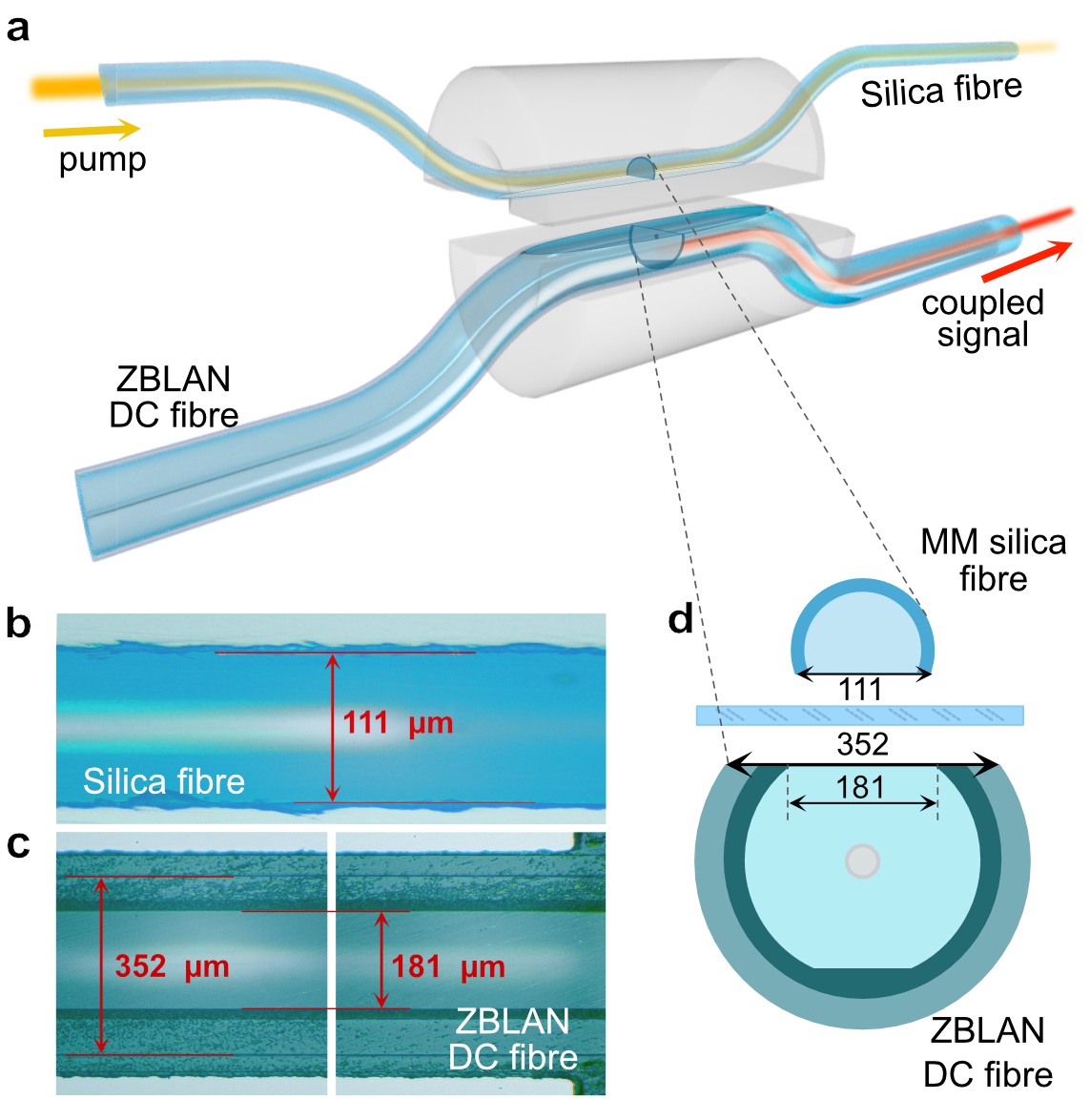}
    \caption{Concept of the pump combiner: \textbf{a} Schematic with the microscope images \textbf{b} and \textbf{c} of polished surfaces of silica and ZBLAN fibres, correspondingly (objective 10X). \textbf{d} characteristic cross-section}
    \label{fig:first}
\end{figure}

Here, we present a new type of hybrid fibre pump combiner that doesn't require thermal post-processing. This combiner uses a pair of side-polished (also known as D-shaped) fibres: multimode silica fibre to deliver the pump radiation and a double-clad ZBLAN fibre. The polishing techniques for both types of fibres are similar and straightforward, and standard commercial polishing pads can be used. We achieved over 80\% coupling efficiency of 980-nm pump signal with the maximum available pump power of approximately 12 W. The deviation of the coupled laser power was as low as 0.09\% during 8 hours of continuous operation. Finally, we demonstrate the performance of the pump combiner in a linear laser cavity based on Er-doped gain fibre, reaching $\sim$15.5\% generation efficiency at around 2780~nm. This pump combiner design is not limited to a specific signal fibre, such as ZBLAN. As we demonstrated, it can employ both passive or active fibres, as well as be translated to fibres of different glass matrices or double-clad fibre geometries. Therefore, this design can be applied to a wide range of fibre laser and amplifier systems and operational wavelengths.

\section*{Results}
\noindent The pump combiner design, presented in Fig.~\ref{fig:first}, was developed via the side-polishing of silica and Er-doped (1mol\% doping concentration) ZBLAN optical fibres, incorporated into house-made 3D-printed polymer holders and assembled into kinematic mounts for optimising the coupling procedure, as described in the Methods section. The width of the fibres polished surfaces, 111 and 352~$\mu$m for silica and fluoride fibres, correspondingly, have been empirically determined during progressive polishing accompanied with coupled pump power monitoring until the optimal conditions have been reached.

\begin{figure}[!t]
    \centering
    \includegraphics[width=0.5\linewidth]{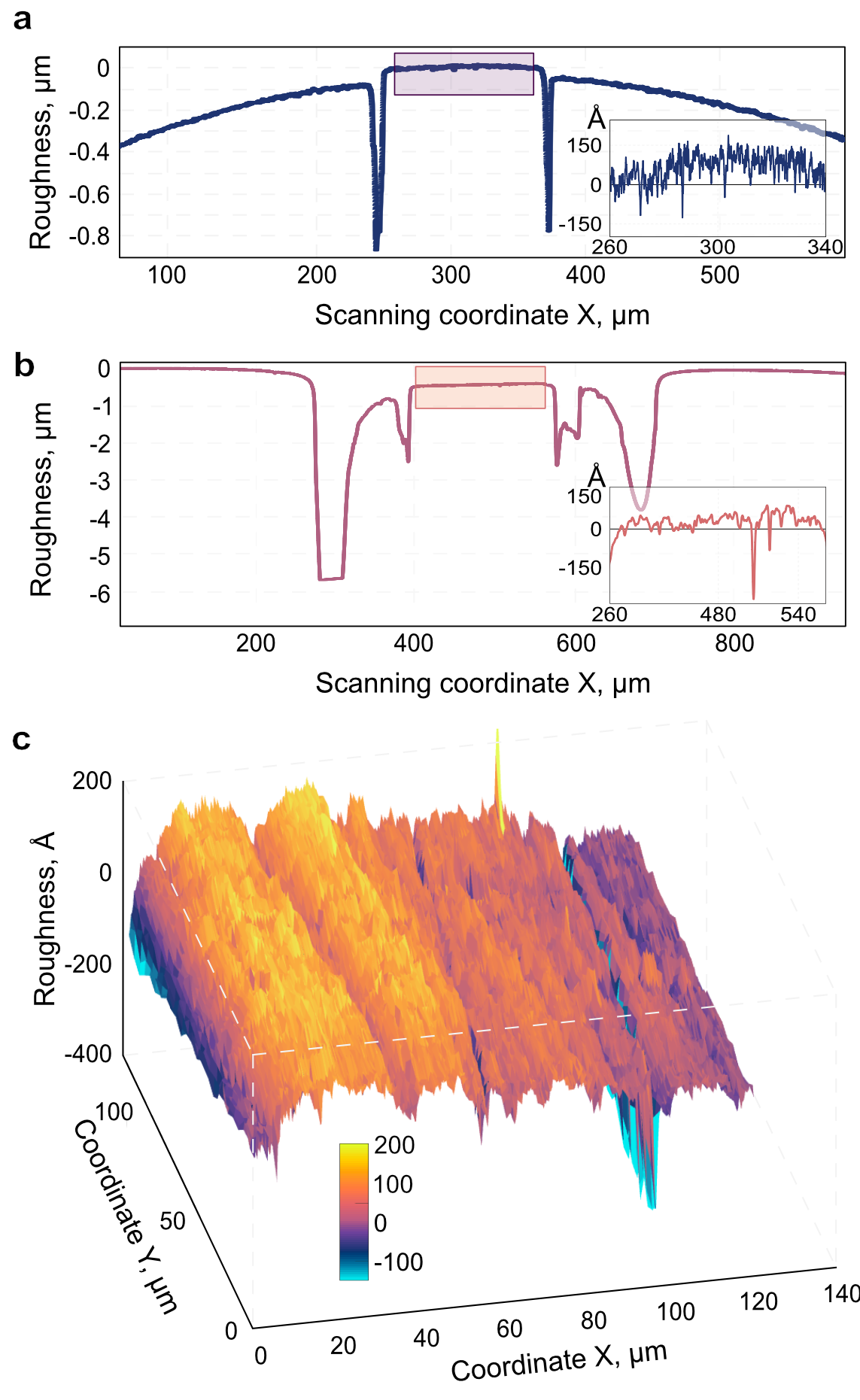}
        \caption{Quality assessment of the polished surfaces of D-shaped fibres. Line-scan of the polished surfaces of \textbf{a} silica and \textbf{b} ZBLAN fibres; \textbf{c} 2D scan of the ZBLAN fibre surface}
        \label{polishing}
\end{figure}

The surface quality presents a decisive factor for the resulting beam combining efficiency. Figures~\ref{fig:first}b and c demonstrate microscope images of the polished surface of the silica and fluoride fibres. The surface profiles across the fibres as a line-scan for both silica and fluoride fibres and detailed 2D scan over ~120$\times$140~$\mu$m$^2$ polished area of ZBLAN fibre are demonstrated in Fig.~\ref{polishing}. The trenches in the profiles correspond to the glue-filled gaps between the fibres and the embedding ceramic ferrules. Due to the different hardness of the optical fibres and ceramic ferrule materials, the profile of polished silica fibre features a small ledge over the ferrule, while fluoride fibre is suppressed. Furthermore, in the case of fluoride fibre, an additional trench can be observed around the first cladding, which is related to the soft low-index resin of the fibre's second cladding. These trenches could be caused by the polishing procedure and measurement approach, which uses the stylus with constant pressure for scanning. Line scans demonstrate the excellent quality of the adjoined surfaces of the fibres, since the grooves left on the glass surface after the polishing procedure are smaller than the wavelength of the injected pump light (980~nm) and should not contribute to scattering losses. A small amount of index-matching oil (1.451 at 980~nm and room temperature) has been applied to fill the gaps between silica and fluoride fibres, restore the second cladding of the latter one, and, overall, improve the coupling of the pump light from silica into fluoride fibre. %The parameters of the index-matching gel and their importance are discussed further. 

\subsection*{Pump combiner efficiency}

\begin{figure}[!bp]
    \centering
    \includegraphics[width=0.5\linewidth]{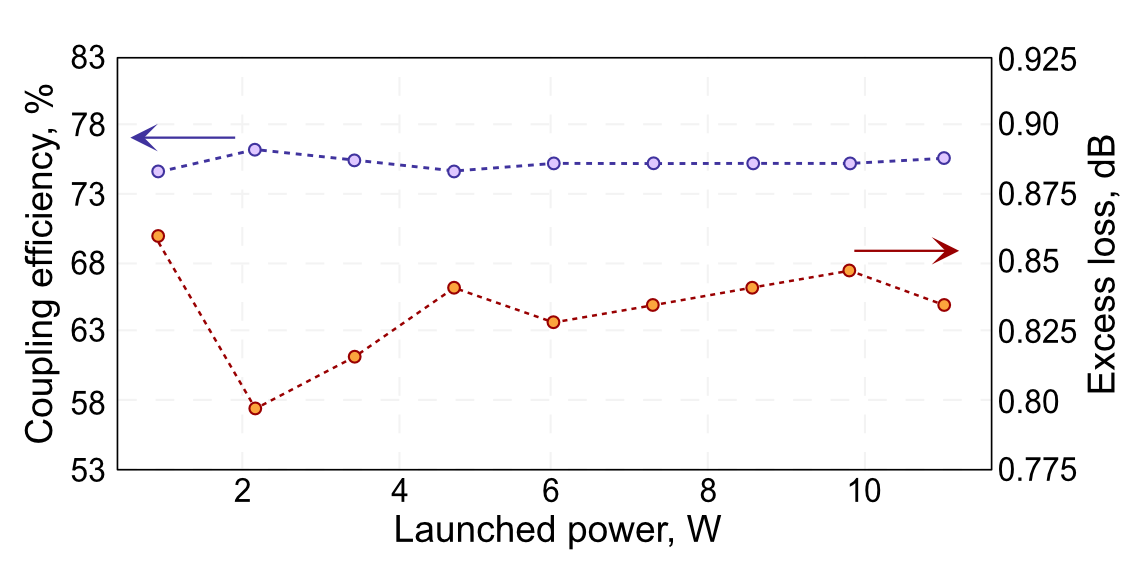}
    \caption{Coupling efficiency and excess loss of the beam combiner at different power.}
    \label{fig:Efficiency}
\end{figure}

For testing the coupling efficiency, a light from a multimode laser diode with a maximum power of 12~W at 980~nm has been launched into the silica fibre. The output power was simultaneously monitored at three arms:  through port (transmission through silica fibre), cross port (coupling into Er-doped ZBLAN fibre), and isolated port (rear end of the fluoride fibre). Figure~\ref{fig:Efficiency} demonstrates the instantaneous coupling efficiency of the pump combiner at different launched powers. The measurements were taken 2-3 minutes after the irradiation was launched. Within the available power range of laser diode stable operation a nearly constant coupling efficiency of 75\% regarding outcoupled power. 
It is worth noting that only negligibly small power could be observed at the isolated port, justifying the directional nature of the pump combiner. Furthermore, within the final arrangement, nearly 66\% of the launched pump power has been outcoupled from silica fibre. The pump light remaining in silica through-port can be recycled or reduced by increasing the length of the polished surface, as the 1-cm length of available ferrules restricted it in the current experiment. Remarkably, the excess loss of the pump combiner (ratio of the launched power from the laser diode to the total output power from all ports) varied from  0.79 to 0.86~dB at 980~nm. As the laser or amplifying system can comprise several pump combiners, used for pumping in either or both co- and counter-propagating directions~\cite{xiao2023numerical}, the light remaining in the silica fibre through the port can be coupled into the active fibre at the later stage.  

\begin{figure}[!t]
        \centering
    \includegraphics[width=0.5\textwidth]{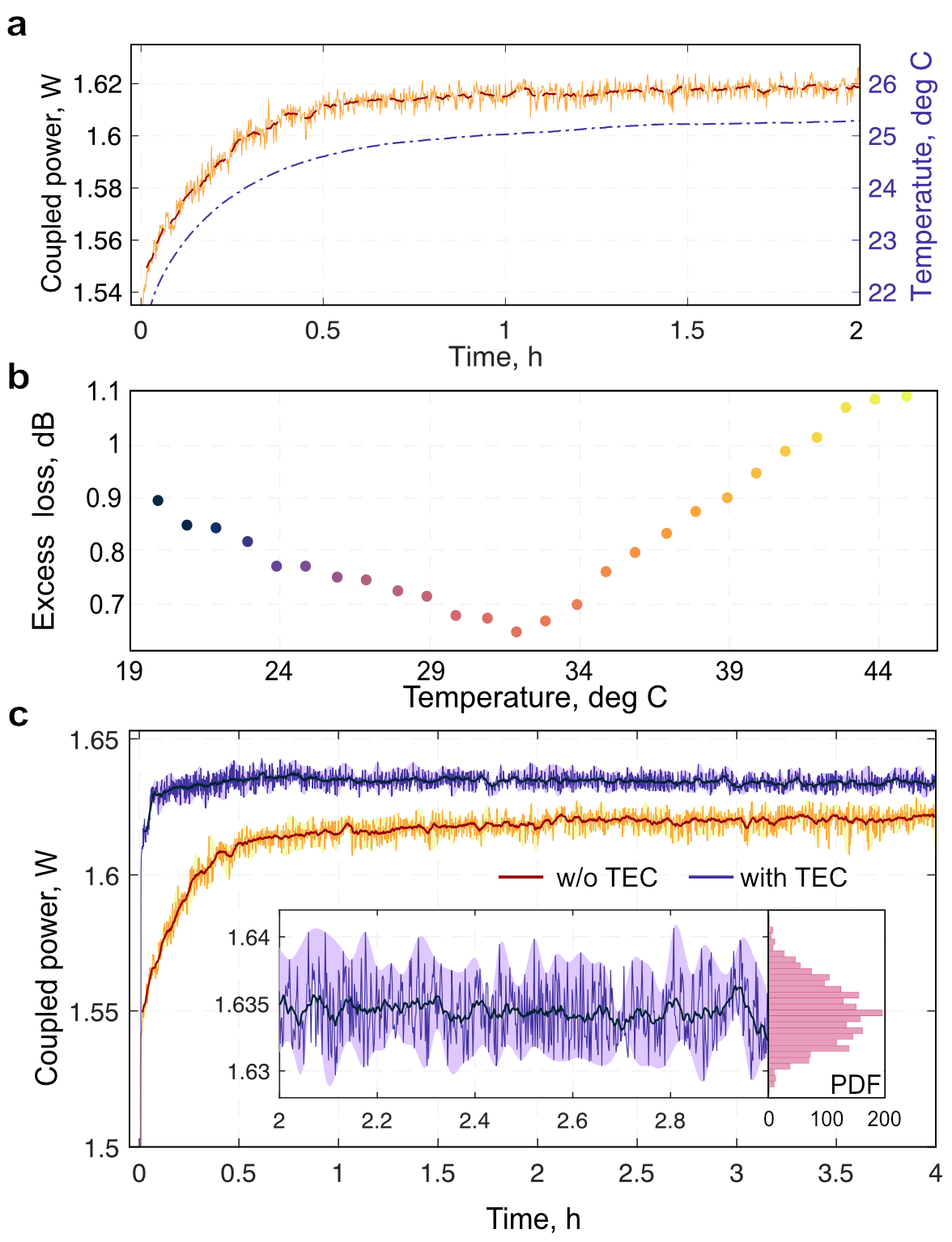}
        \caption{Long-term stability of the coupled power. \textbf{a} Build-up of the coupled power without thermal stabilisation; \textbf{b} Variation of pump combiner excess losses at different temperatures; \textbf{c} Comparison of the stability of the coupled power with and without thermal stabilisation over 4 hours of continuous operation. Inset: Stable coupling with RMS of 0.09\% depicting power probability distribution.}
        \label{stability}
\end{figure}

Figure~\ref{stability} demonstrates an example of the long-term operation of the pump combiner. Here, the pump diode was set to operate at an intermediate power of 3.5~W. 
The build-up dynamics of the coupled power when the pump combiner was not stabilised against thermal effects, particularly those related to laser irradiation, is demonstrated in Fig.~\ref{stability}(a). As soon as the laser diode is switched on, the power from the through port almost immediately reaches a level of 95\% of the average. Over the next 20 minutes, the power rose to a level of 99\% and then, after only two hours of running time, got entirely stabilised at an  root mean square (RMS) value of 1.62~W. Starting from this point, the next 6 hours of continuous monitoring showed the deviation of the RMS of average power coupled into ZBLAN fibre to be as low as 0.12\%. The RMS deviation value of less than 0.2\%  is a reasonable value for fibre lasers. It is primarily attributed to the noise of the diode itself and the noise of the detector. At the end of the stability check, the pump combiner was visually investigated using a microscope, which did not reveal any signs of side-polished fibre degradation. 

\subsection*{Thermal stabilisation}
Thermal effects inside the pump combiner present a crucial point on the way to increasing the coupling efficiency \cite{magnan2020fuseless}. Indeed, the simultaneous measurement of the temperature of the pump combiner configuration during the stability check confirmed that its temperature rises from the room temperature, of 21~$^\circ$C, to up to 25.5~~$^\circ$C during stabilisation of the coupling efficiency. Importantly, the temperature increase follows a similar trend as the output power from the ZBLAN fibre cross port, as shown in Fig.~\ref{stability}(a).

The correlation between the coupling efficiency and the internal temperature of the combiner design was analysed in the range from 20 to 45~$^\circ$C (Fig.~\ref{stability}b). For this, a thermoelectric cooler system based on Peltier elements (TEC) with temperature sensors for feedback has been assembled. The experimental results for the same launched power level of 3.5~W at 980~nm confirm that heat dissipation is an important effect for the combiner efficiency. Thus, the optimal operation temperature is 32~$^\circ$C when the excess loss drops from 0.84 below 0.65~dB. At the same time, pump coupling efficiency increases to 81.5\%. We attribute this to the thermal-dependant refractive index of the index-matching gel. %, shown in Fig.~\ref{stability_w_TEC} (a). 
Furthermore, the significant thermal expansion of ZBLAN glass may contribute to improved coupling efficiency, as the polished surface of the expanded ZBLAN fibre achieves better physical contact with the side of the silica fibre.~\cite {zhu2010high}. 

Next, a long-term stability test was conducted for the temperature-stabilised beam combiner. In this case, to enhance the heat conductivity and temperature stability of fibre holders, metallic fibre ferrule holders were used instead of the 3D-printed polymer ones. Figure~\ref{stability}(c) compares the stabilisation processes of the output power from ZBLAN fibre at 3.5W pump power, with and without an active temperature control system in the pump combiner. The result of the thermal stabilisation is the almost immediate output of the laser build-up to 98.5\% of the maximum average power. After 20 minutes, the power has been fully stabilised. The deviation of the output power RMS has been measured analogously from the second hour of combiner operation (see Inset in Fig.~\ref{stability}b). With active temperature stabilisation, the RMS stability deviation slightly decreased to 0.09\%.

%\begin{figure}[!bp]
 %       \centering
%   \includegraphics[width=0.95\textwidth]{Figures/Temp sabilisation.png}
   %     \caption{Stability of the coupled power with and without thermal stabilisation over 4 hours of continuous operation. Inset: (b) Stable coupling with RMS of 0.09\% depicting power PDF.}
  %      \label{stability_w_TEC}
%\end{figure}

\subsection*{Performance in a fibre laser cavity}
\begin{figure}[!]
    \centering
    \includegraphics[width=0.5\linewidth]{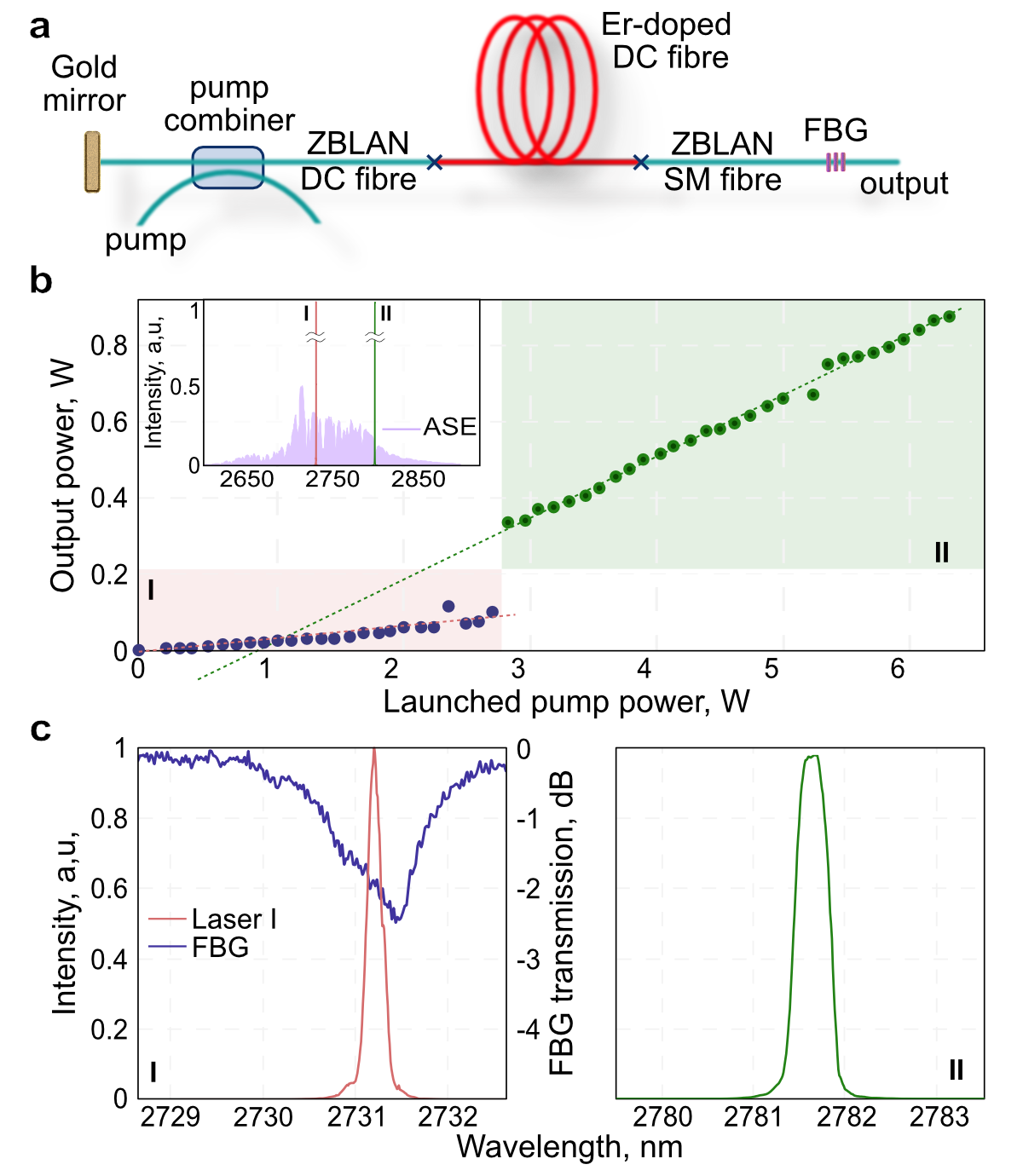}
    \caption{Performance of the pump combiner in a laser system. \textbf{a} Laser system setup; \textbf{b} Laser efficiency. Inset: ASE and laser spectra at areas I and II; \textbf{c} Output spectra at different pump powers. }
    \label{fig:Laser}
\end{figure}

Finally, we validated the performance of the demonstrated pump combiner in the laser cavity. Due to the directional nature of the pump combiner, the rear section of the Er-doped ZBLAN fibre in the initial design was left unpumped, leading to potential parasitic losses when integrated into the laser cavity. Consequently, an equivalent pump combiner was developed using both passive silica and ZBLAN double-clad fibres. Notably, the passive ZBLAN fibre shares similar optical properties with the previously used Er-doped ZBLAN fibre, allowing the pump combiner built in all-passive fibre configuration to achieve comparable pump coupling efficiency as discussed in earlier sections.Furthermore, the passive ZBLAN fibre features a circular cross-section of the first cladding, which lifted the requirement on its precise orientation in the ferrule for polishing, as was in the case of Er-doped fibre with double-D-shaped first cladding (as shown in Fig.~\ref{fig:first}d). The pump combiner was spliced to one end of 7-mol\% Er-doped fibre, as shown in Fig.~\ref{fig:Laser}(a). The other end of the active fibre was spliced with a single-mode step-index passive ZBLAN fibre. The laser cavity was formed by a gold mirror, connected via an FC/PC fibre adapter to a beam combiner rear ZBLAN port and a fibre Bragg grating (FBG). \textit{Methods} section presents a detailed description of the laser system. Figure.~\ref{fig:Laser}(c) depicts the transmission spectrum of the FBG.

As shown in Fig.~\ref{fig:Laser}(b), the laser threshold of continuous-wave operation at around 2.73~$\mu$m central wavelength corresponds to $\sim$0.4~W pump power at 980~nm. The maximum output power at 2.73~$\mu$m reached almost 100~mW when the pump power was set at approx.~2.9~W (an area I marked pale red). Here, the laser efficiency is only 3.6\%. As the maximum of the emission spectrum in the optical fibres highly doped with Er$^{3+}$ ions corresponds to $\sim$2.8~$\mu$m~\cite{gan2021high}, further pump power increase led towards the shift of the laser generation to longer wavelengths, as shown in the inset of Fig.~\ref{fig:Laser}(b) alongside with the amplified spontaneous emission (ASE) spectrum. In this case, the flat polished fibre tip operated as a cavity mirror, replacing the FBG. With the increase of the launched into ZBLAN pump power to almost 6.5~W, the output power reached 870~mW at 2.78~$\mu$m (area II marked green), which corresponds to 15.5\% slope efficiency. The primary factors limiting the efficiency and laser output power are the splice losses of dissimilar ZBLAN fibres and pump power dissipation at the stripped splice region of double-clad active and passive fibres. 

Figure.~\ref{fig:Laser}(c) depicts the detailed output spectra for both cases. The generation at Near-IR wavelength range around 1550~nm, characteristic for Er-doped optical fibres, was not pronounced. This is a characteristic feature of fibres with a high concentration of Er$^{3+}$ ions. The reduction of the distance between adjacent rare earth ions enhances the impact of energy-transfer upconversion from lower $^4$I$_{13/2}$ laser level and allows for energy recycling to the upper $^4$I$_{11/2}$ laser level \cite{pollnau2002energy,jackson2022spectroscopy}. This has been confirmed by negligible output power recorded after a short pass filter, cutting out wavelengths beyond 2.5~$\mu$m.

\section*{Discussion}
\noindent
This work demonstrates a pump combiner designed to launch the light of a silica-fibre pigtailed pump laser diode into a fluoride-based fibre laser. The proposed design uses a pair of side-polished (D-shaped) fibres. With the help of high-precision actuators, refractive index matching and thermal stabilization, almost 80\% coupling efficiency has been achieved at 980~nm. The overall excess loss is below the 0.65-dB level. By optimising the side-polished fibre geometries, such as polishing depth and area and the orientation of asymmetrical signal-guiding double-clad fibres (as in the case of the double D-shaped profile of the first cladding), the performance of the design can be further improved. The study of temperature regimes provides the possibility of using the pump combiner in active mode, allowing precise regulation of the power injected into the laser resonator and stabilisation of the output power at the required level. The demonstrated Er-doped fibre laser performance at around 2.8~$\mu$m, with 870~mW average power in continuous wave operation and 15.5\% efficiency, justifies the feasibility of the proposed concept of a pump combiner to be successfully integrated into various cavity configurations to create all-fibre lasers. 

\begin{figure}[!]
    \centering
    \includegraphics[width=0.5\linewidth]{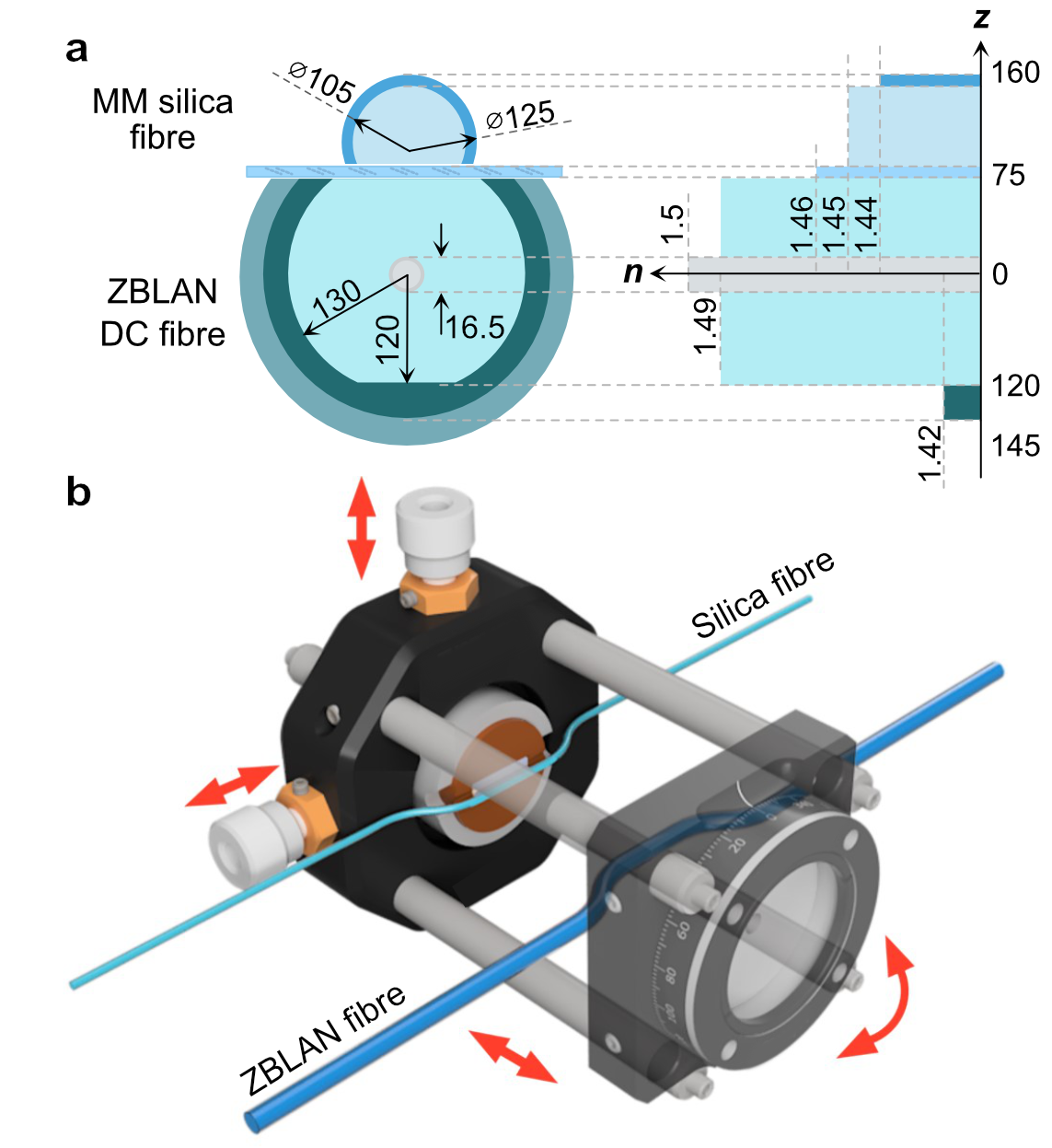}
    \caption{Fibre arrangement in the beam combiner. \textbf{a} Cross section of beam combiner with the corresponding schematic of refractive index profiles of the used fibres. \textbf{b} Schematic of the side-polished fibre alignment for optimising coupling efficiency. }
 \label{fig:schematic}
\end{figure}

 Notably, the pump combiner operates solely on the principle of evanescence field coupling, which does not impose any spectral limitations compared to wavelength-division multiplexors. The geometry of side-polished fibres can be applied both for a specific pump wavelength or a combination of wavelengths, e.g. in the case of dual-pumping~\cite{bawden2021ultrafast}. Furthermore, it can be used in laser and amplifier systems employing co- and counter-propagating pumping schemes. Moreover, it is essential to highlight that such pump combiner can comprise other fibres that are challenging to splice with pump-delivering silica fibres, e.g. chalcogenide or polymer fibres. However, one of the apparent limitations of the suggested design of a pump combiner is the limited ability to incorporate multiple pump ports to be used with several pump sources in high-power laser systems. Nevertheless, the demonstrated pump combiner concept effectively resolves the principal challenge, enabling the compatibility of silica and fluoride-based fibres. In the case of designing high-power laser systems, there are no constraints on implementing a commercially available silica-based multiple-pump-port fibre combiner at the silica pump-delivering port.

We envision that the design presented in this work is highly versatile and will become an enabling attribute for developing low-loss tuneable fluoride-based optical couplers. With the adequate side-pumping technique, such developments would produce simple, flexible and robust all-fibre laser systems operating at the mid-IR wavelength range.

\section*{Materials and methods}\label{Methods}

\subsection*{Fibres and pump combiner configuration}

To validate the concept of a pump combiner based on side-polished fibres, we initially used a double-clad ZBLAN fibre with double D-shaped geometry of the fist cladding, doped with 1mol\% of Er$^{3+}$ ions (from Le Verre Fluor{\'e}), as a signal port. The core diameter was 16.5$\mu$m; the first cladding had dimensions of 240 by 260$\mu$m, and the second cladding diameter was 290~$\mu$m. Figure~\ref{fig:schematic}(a) schematically shows the corresponding refractive index profiles of the fibres, measured using IFA-100 Multiwavelength Optical Fiber Analyzer at 633~nm and extrapolated to the pump wavelength of 980~nm~\cite{boilard2022probing}. The core and first cladding numerical apertures of ZBLAN fibre were 0.12 and 0.4, correspondingly. 
The pump delivery silica fibre had 105 and 125 core/cladding diameters, respectively. 

The fibres were glued into the ceramic ferrules and fixed into 3D-printed polymer (Polylactic Acid) holders. The holders were designed to ensure compatibility with the commercially available mounts for $\varnothing$ 1-inch round optics (Fig.~\ref{fig:schematic}b).

The polishing process was performed using a home-built machine based on the rotating platform and involving lapping films with a grit size as low as 0.03 $\mu$m. The polishing depth of fluoride fibres was around 50~$\mu$m from the surface of the second cladding. Figure~\ref{fig:first}(b) shows a microscope image of the polished ZBLAN fibre, depicting the side polished ferrule, the width of the overall surface (352~$\mu$m), and the width of the first cladding of 181~$\mu$m. The rest of the ferrule bore with 440~$\mu$m diameter has been filed with the two-component epoxy glue with optical properties matching the coating of ZBLAN fibre. The silica fibre has been polished down to a residual height of 24~$\mu$m regarding the fibre centre point (Fig.~\ref{fig:first}b). As a result, the fibres had $\sim$110~$\mu$m of direct contact width (Fig.~\ref{fig:first}d) over the entire length of the ceramic ferrule, i.e. $\sim$1~cm. The high quality of the polished surface was assessed in a line scan across the fibre surface and a 2D scan over ~120$\times$140~$\mu$m$^2$ polished area by using the surface profiler, Benchtop Stylus Surface Profiler P-7, from KLA-Tencor.

Once the polishing process reached a suitable surface quality to avoid light scattering, the parts of the pump combiner were placed into the mounts. The two-adjuster kinematic mount on one side aligned fibres along the X and Y axes. The counterpart high-precision rotation mount enabled relative angular adjustment of the fibres (Fig.~\ref{fig:schematic} b). 

After characterising the pump combiner efficiency, the analogous design has been assembled using passive double-clad ZBLAN fibre. Passive fibre has a core diameter of 14~$\mu$m and a 250-$\mu$m circular first cladding. The core and first cladding numerical apertures of passive ZBLAN fibre were 0.12 and 0.4, respectively. As the active and passive fibres have similar waveguiding parameters, the overall beam coupling efficiency was the same as in the demonstrated pump combiner comprising 1mol\% Er-doped fibre.

\subsection*{Measurement and characterisation}
A laser diode with a maximum power of 12~W at 980nm is pigtailed with multimode silica fibre, and the output power has been simultaneously monitored at three arms of the pump combiner:  through port (transmission), cross port (coupling) and isolated port. We used power meter consoles with thermal power sensors (S425C-L, from Thorlabs). All fibre ports have been terminated with temporary FC/PC connectors without end cap protection.

\subsection*{Laser Components Specification}
 The laser comprises a 2.9-m long section of a double-D-shape double-clad ZBLAN fibre, doped with 7~mol\% Er$^{3+}$ ions (Le Verre Fluor\'e). The rare-earth-doped fibre has a core, cladding, and second cladding diameters of 16.5, 240*260 and 290~$\mu$m, correspondingly. The numerical apertures of the core and first cladding are 0.12 and 0.46, respectively. 
 The pump light at 980~nm was launched into the active fibre through the developed pump combiner. While the maximum pump power of the laser diode was 12 W, about 34\% of the available pump power remains in the silica fibre through-port. Accounting further for the excess loss of the pump combiner of ~0.65 dB, the actual pump power launched into the ZBLAN fibre is $\sim$6.5~W. 
 The single-mode passive fibre (from FiberLabs) has a 3.6-$\mu$m core diameter and numerical aperture of 0.26, which ensured good agreement of mode field diameters at the laser operation wavelength. The fibres have been fusion-spliced to form the all-fibre laser cavity using Vytran GPX3400.
 
 The FBG has been inscribed using a phase mask-based Talbot interferometer and an infrared regeneratively amplified Ti:Sapphire femtosecond laser, operating at the second harmonic wavelength of 400~nm~\cite{chiamenti2021first}. The FBG features 2.731-$\mu$m Bragg wavelength and $\sim$40\% reflectivity.

\section{Acknowledgement}
    \noindent
The work has been funded by the German Federal Ministry of Education and Research (BMBF) and supervised by the VDI Technology Center under number 13N15464 "Leibniz Center for Photonics in Infection Research (LPI): Multidimensional, multimodal, intelligent imaging platforms". The authors also acknowledge the support from the European Regional Development Fund. The Authors thank Le Verre Fluor\'e for providing fibre samples and Annett Gawlik for assessing the surface quality of polished fibres. 
The Authors appreciate fruitful discussions with colleagues from the Specialty Fibre Competence Center at Leibniz Institute of Photonic Technologies. %The authors acknowledge Annett Gawlik for measuring the polishing quality in Fig.~\ref{polishing}.

%\begin{figure*}[!b]
%    \begin{adjustbox}{minipage=\linewidth-4pt,rndframe={width=1pt,color=aa}{5pt}}
%        \centering
%        \includegraphics[width=0.95\linewidth]{Figures/polishing.png}
%        \caption{Quality assessment of the polished surfaces of D-shaped fibres. \textbf{a} Microscopic pictures of silica and \textbf{b} ZBLAN fibre surface (objective 20X); \textbf{c} Line-scan of the polished ZBLAN fibre surface roughness; \textbf{d} 2D scan of the ZBLAN fibre surface}
%        \label{polishing}
 %   \end{adjustbox}
%\end{figure*}

\section{Aauthor contributions}
    \noindent
B.P. developed a homemade polishing setup and performed all experiments testing the pump combiner. K.G. inscribed FBG. U.H. and his group fabricated tailored phase masks for FBG inscription. B.P. and K.G. performed a laser experiment. M.C. conceived the idea and led the project. B.P. and M.C. wrote the manuscript with the involvement of all authors.
\noindent
Experimental Data underlying the results presented in this paper are available in Ref.\cite{data}.

\section{Conflict of interest}
    \noindent
    The authors declare no conflicts of interest.

\bibliography{ref}
\bibliographystyle{abbrv}

%\nocite{*}

%\end{document}

\end{document}